# PATTERN DETECTION ON GLIOBLASTOMA'S WADDINGTON LANDSCAPE VIA GENERATIVE ADVERSARIAL NETWORKS

[1]Abicumaran Uthamacumaran*

[1]Dept. Of Physics, Concordia University, Montreal, QC, Canada.

Correspondence: a_utham@live.concordia.ca

**ABSTRACT**

Glioblastoma (GBM) is a highly morbid and lethal disease with poor prognosis. Their emergent properties such as cellular heterogeneity, therapy resistance, and self-renewal are largely attributed to the interactions between a subset of their population known as glioblastoma-derived stem cells (GSCs) and their microenvironment. Identifying causal patterns in the developmental trajectories between GSCs and the mature, well-differentiated GBM phenotypes remains a challenging problem in oncology. The paper presents a blueprint of complex systems approaches to infer attractor dynamics from the single-cell gene expression datasets of pediatric GBM and adult GSCs. These algorithms include Waddington landscape reconstruction, GANs (Generative Adversarial Networks), and fractal dimension analysis. Here I show, a Rössler-like strange attractor with a fractal dimension of roughly 1.7 emerged in all n = 12 patients' GAN-reconstructed patterns. The findings suggest a *strange attractor* may be driving the complex dynamics and adaptive behaviors of GBM in signaling state-space.

## 1. INTRODUCTION

Cancer remains amidst the globally leading cause of disease-related death (World Health Organization, 2021). Glioblastoma (GBM) is a lethal disease, and the deadliest of brain cancers affecting both children and adults. The poor prognosis and hallmark of cellular heterogeneity in GBM is attributed to the signaling dynamics between a subpopulation of therapy- resistant cells, referred to as glioblastoma stem cells (GSCs), and their tumor microenvironment (Jung et al., 2019). GSCs possess the ability to self-renew and maintain the distinct cellular phenotypes of the tumor (Jung et al., 2019).

Outstanding questions in GBM progression include the mosaicism of GBM/GSC populations, the ability of GBM cells to transition in and out of the GSC state and identifying the molecular networks steering their cell fate decisions via the interactions with their tumor microenvironment (niches) (Mitchell et al., 2020). Furthermore, there remains no effective method to predict how clinical therapies may alter GBM cell states, or potentially increase (or decrease) their GSC stemness (i.e., transition to GSC state). Further, recent studies demonstrate current therapies may enhance GSC stemness and hence, promote emergent behaviors such as GBM aggressivity and adaptiveness (Xiong et al., 2019). These are fundamental questions needed to be addressed to understand how the GBM cell state evolves in time and thereby predict its therapeutic response to treatments (Mitchell et al., 2020). Within the language of dynamical systems, these clinically - relevant questions inquire upon the *attractors* to which the GBM/GSC cell fates evolve towards in the signaling state-space (Huang et al., 2009; Jia et al., 2017). As demonstrated in the present study, *complex systems theory* may be able to provide key insights to these fundamental questions by providing novel tools to infer the complex dynamics and cybernetics of GBM.

Complex systems theory, or simply complexity science, is a recent paradigm in physics devoted to the study of *complex systems* (Bossomaier and Green, 2000). Cancers are such complex systems (Uthamacumaran, 2021). That is, their emergent properties, patterns, and signaling networks steering their cell fate dynamics (i.e., cybernetics) comprise of an *irreducible* system. Complex systems are systems composed of many interacting parts, the undivided whole of which gives rise to emergent behaviors (Bossomaier and Green, 2000). In biological cybernetics, complex systems are often networks of complex processes such as protein interactions, gene regulatory relationships, cancer-immune dynamics, etc. In Aristotelian terms, complex systems are systems in which the *whole is greater than the sum of its parts*, indicating the nonlinear interactions between the parts and its environment (i.e., interconnectedness) (Bossomaier and Green, 2000; Shalizi, 2006). Think of the self-organization of beehives, stigmergy in ant colonies, stock market fluctuations, traffic flow, cybernetics of social networks, patterns of fluid turbulence, physiological oscillations, and cellular gene expression dynamics; these are some of the many examples of *complex systems* (Wolfram, 1988; Goldberger, 2006; Ladyman and Wiesner, 2020). However, the paradigm of complex systems has shown that even simple algorithms or computer programs can be as complex as any naturally occurring complex system and can simulate/model any complex system (i.e., the Church-Turing thesis and Turing's universality). Examples of this includes Conway's game of life and elementary cellular automata (Wolfram, 1988). Therefore, complexity science provides a general framework to study difficult many-body problems in physics within the framework of algorithms and computational physics.

Some general characteristics observed in complex systems include emergence, nonlinear dynamics, spontaneous self-organization, adaptive behaviors, collective dynamics, multi-layered networks, multi-nested feedback loops, recursion, fractal structures, pattern formation, computational irreducibility, multi-scale information flow, criticality, long-term unpredictability, non-locality, and undecidability (Wolfram, 1988; Ladyman and Wiesner, 2020). To overcome the inadequacy of traditional analytic approaches in deciphering the patterns and emergent behaviors of complex systems, complexity science advocates the marriage between data science (machine intelligence) and interdisciplinary tools from various branches of physics including dynamical systems theory, graph theory/networks, information theory, and statistical physics (Shalizi, 2006; Thurner et al., 2018). Herein, some of these complex systems tools are assessed for causal pattern analysis in cancer (GBM) gene expression.

A major obstacle in the advancement of clinical oncology remains the lack of time-series gene expression datasets. This presents a fundamental barrier in studying the complex dynamics of cancer signaling and their attractor reconstruction. Even if present, these are complex datasets with many interacting variables and high technical noise from the single-cell RNA-Sequencing (scRNA-Seq) technologies (Kim et al., 2015). Denoising and data filtering algorithms exist to partially overcome these challenges. In dynamical systems theory, attractors are the self-organized causal patterns to which the trajectories of complex systems (processes) are bound to in state-space reconstruction ((Janson, 2012). Attractors govern the stability, regularity, and long-term control predictability of the complex system's behaviors (Strogatz, 2015). Without time-series analysis, the dynamics of GBM cell fates are limited to simple stable attractors in phase-space, such as fixed-points or limit cycles (periodic oscillations) (Janson, 2012; Strogatz, 2015). How can we detect the presence of more complex attractors in GBM's signaling state-space without time-series datasets?

As a solution to this problem, herein, a three-step blueprint from complex systems tools is provided for detecting causal patterns (attractors) in the gene expression state-space of GBM/GSC cells and thereby infer GBM cell fate dynamics (developmental trajectories): (a) Waddington landscape reconstruction from the scRNA-Seq count matrices, followed by (b) Deep learning-based pattern generation, and (c) fractal dimension analysis of the machine-generated patterns. Deep learning networks, arguably the most advanced of classical machine learning algorithms, provide robust tools for pattern detection in complex diseases (Esteva et al., 2019; Topol, 2019). Despite the existence of Deep learning algorithms, their use in the detection of causal patterns in cancer gene expression remains under-applied. Here, I demonstrate the application of Generative Adversarial Networks (GANs), a type of Deep learning architecture, to identify causal patterns in the GBM cell fate landscapes. The findings show the investigated complex systems approaches provide a computationally efficient and cheap alternative for causal pattern discovery from single-cell gene expression datasets *without* time-series measurements.

2. METHODS

   2.1. **Single-cell Datasets:** The scRNA-Seq gene expression matrices for pediatric GBM and adult GSC were obtained from the SingleCell Portal repositories from Neftel et al., 2019 and Richards et al., 2021, respectively. GBM patient samples from Neftel et al. contained the scRNA-Seq counts of four distinct phenotypes (cellular states): macrophages, malignant GBM cells, oligodendrocytes, and T-cells. Adult GSC samples consisted only of stem cells (identified by specific cell surface markers). Twelve random samples of scRNA-Seq expression count matrices were selected consisting of n = 6 pediatric GBM patient samples from the Neftel et al. study, and n = 6 adult GSC samples from the Richards et al. study. A description of the experimental protocol applied to derive these single-cell datasets is provided in their respective citations.

   2.2. **scEpath Algorithm:** Waddington landscape reconstruction is a tool to study complex processes within complex adaptive systems, such as the cell fate decisions in populations of GBM/GSC cells (Uthamacumaran, 2021). The Waddington energy landscape is a metaphor to describe the cell fate developmental trajectories of stem cells (Waddington, 1957). Stem cells occupy higher energy states, represented by the local energy maxima (peaks and hills) of the energy landscape, while the differentiated (mature) cell fates are represented by the local energy minima (low energy states) corresponding to valleys of the landscape (Waddington, 1957). The Waddington landscape thus provides a three-dimensional visualization of cell fate choices and their emergent patterns during cell fate transitions. The Waddington energy landscape can be treated as a representation of the cells' gene expression (signaling) state-space.

Waddington landscape reconstruction was performed using the scEpath MATLAB-package (Jin et al., 2018). The **S**ingle **c**ell **E**nergy path (scEpath) is a trajectory inference algorithm comprised of various tools from statistical physics and data science, for mapping the attractors (cell fate trajectories) underlying cell fate decisions. It reconstructs the 3D-energy landscape of cells and infers regulatory relationships from their transcriptional dynamics (Jin et al., 2018). To perform scEpath analysis on our data, the log-normalized count matrices (with respect to their gene expression values) were first pre-processed by filtering out zero counts. The scEpath MATLAB code from Jin et al. (2018) was then assessed on the gene expression matrices of the 12 randomly selected patient samples. Each pediatric GBM sample consisted of roughly 100-200 cells, and >20,000 genes, whereas each adult GSC sample consisted of roughly 500 cells, and >19,000 genes.

To reconstruct the 3D Waddington energy landscapes, scEpath performs principal component analysis (PCA) on the energy matrix $E = (E_{ij})$ obtained from the gene expression counts and fits a potential energy surface using piecewise linear interpolation over the first two PCA components and the single cell energy (scEnergy) of each cell. Cells are then colored according to unsupervised clustering which groups cells with similar gene expression patterns (transcriptional states). The cell size is proportional to scEnergy. The energy of each cell state (scEnergy), $E_j$, on the Waddington landscape is computed according to the following function:

$$E_j(y) = \sum_{i=1}^{n} E_{ij}(y) = -\sum_{i=1}^{n} y_{ij} \ln \frac{y_{ij}}{\sum_{k \in N_i} y_{kj}}, (1)$$

where $y_{ij}$ represents the normalized gene expression level (between 0 and 1) of gene i and cell j, and $N(i)$ is the neighborhood of node-i in the network. Each gene is assigned a local energy state $E_{ij}$ (Jin et al., 2018). The scEnergy is then combined with a distance-based measure and structural clustering to reconstruct the 3D energy landscapes. The cell-state on the scEpath Waddington landscape corresponds to which discrete bin its mRNA levels fluctuate within (Jin et al., 2018). The cell state distribution on the scEpath landscape can be defined as attractors, a term borrowed from dynamical systems theory used to describe a causal pattern to which the cell fate dynamics are bound to. To infer cell lineages from these cell states, scEpath constructs a probabilistic directed graph in which nodes represent cell clusters, and the edges are weighted by cell state transition probabilities. The cell state transition probability such that a cell state from one cellular cluster with a particular scEnergy (i.e., attractor) can transition to another nearby attractor was calculated using the Boltzmann–Gibbs distribution by the scEpath algorithm (Jin et al., 2018).

- **2.3. GAN Lab:** GAN is a type of Deep learning Neural network, a subset of artificial neural networks (ANN), with multiple layers between the input and output layers. It is a powerful unsupervised machine learning algorithm resembling the neural architecture of the human brain, used for pattern recognition and detection from large, complex datasets (Gonog and Zhou, 2019). The basic structure of the GAN consists of two multilayered ANNs: a generator, and a discriminator. The two networks compete to optimize the pattern generation (Gonog and Zhou, 2019). The generator learns the mapping from a latent space to data distribution (manifold), closely resembling the original data distribution (i.e., cell fate patterns on the scEpath landscape). The discriminator attempts to distinguish between the input (real) dataset and the new pattern generated from the generator. The goal of the generative network is to trick the discriminator into thinking that the novel data distribution it produced is from the input data distribution (Kahng et al., 2019). Hence, GANs can autonomously learn the regularities and patterns of an input dataset, from which complex patterns can be generated. Herein, GAN Lab was used to generate possible attractor patterns from the GBM/GSC gene expression datasets. GAN Lab uses TensorFlow.js, an in-browser GPU-accelerated Deep learning library implemented in Javascript (Kahng et al., 2019).

The scEpath landscape patterns (the distribution of cell states) were carefully traced onto the GAN Lab canvas. Hyperparameters were tuned for optimal learning such that the learning rate using a Stochastic Gradient Descent (SGD) optimizer was 0.01 for the generator and 0.1 for the discriminator, with 99 artificial neurons and a single hidden layer for both. Gaussian noise was present during all GAN-training and pattern generation. The generator was tuned to 1 update per epoch, while the discriminator was set to 2 updates per epoch, allowing the discriminator's performance to be optimal. The GAN converged to patterns after many training iterations (> 5000 epochs) when the maximal amount of purple grid cells were observed in the background of the GAN-interface, indicating the highest classification of a new pattern generated (See Figure 2H).

- **2.4. Fractal Analysis:** The fractal dimension is a non-integer dimension, FD. A fractal dimension suggests the existence of a scaling law describing the complexity and roughness (self-similarity) of the pattern (Mandelbrot, 1982). A fractal dimension is also a characteristic signature of *deterministic chaos* in dynamical systems (Frederickson et al., 1983). Strange attractors, the causal patterns to which the trajectories of a chaotic system are bound to, occupy fractal dimensions in phase-space (Ruelle, 1980). The ImageJ plugin FracLac was used to compute the fractal dimension (FD) of the GAN-generated patterns. The images of the GAN-reconstructed attractors were enhanced in saturation to 25% and converted to binarized images on the ImageJ Plugin. FracLac then performs FD calculation via the Box-counting algorithm. The description of the Box-counting algorithm is as follows: Let L be the line length, $\varepsilon$ be the box size, and $N(\varepsilon)$ be the number of boxes which can divide the pattern/object into self-similar substructures. The slope of the log-log plot of N and $\varepsilon$, if it exists, provides the box-count fractal dimension of the object/pattern, as given by:

$$FD = \lim_{\varepsilon \to 0} \frac{lnN(\varepsilon)}{\ln\left(\frac{1}{\varepsilon}\right)} \quad (2)$$

All other statistical analysis was performed using GraphPad Prism 8.4.3.

## 3. RESULTS

**3.1. *scEpath algorithm reconstructs the cell fate attractors underlying GBM and GSC gene expression dynamics.*** A bifurcation is always seen from high energy cell states to low energy cell states on the landscapes. Bifurcations are the precursors for the emergence of complex attractors in the phase space of dynamical processes (Strogatz, 2015). Cell clusters are determined by gene expression similarity (see methods section). The scEpath algorithm infers a flow structure in cell states suggesting the presence of a global attractor.

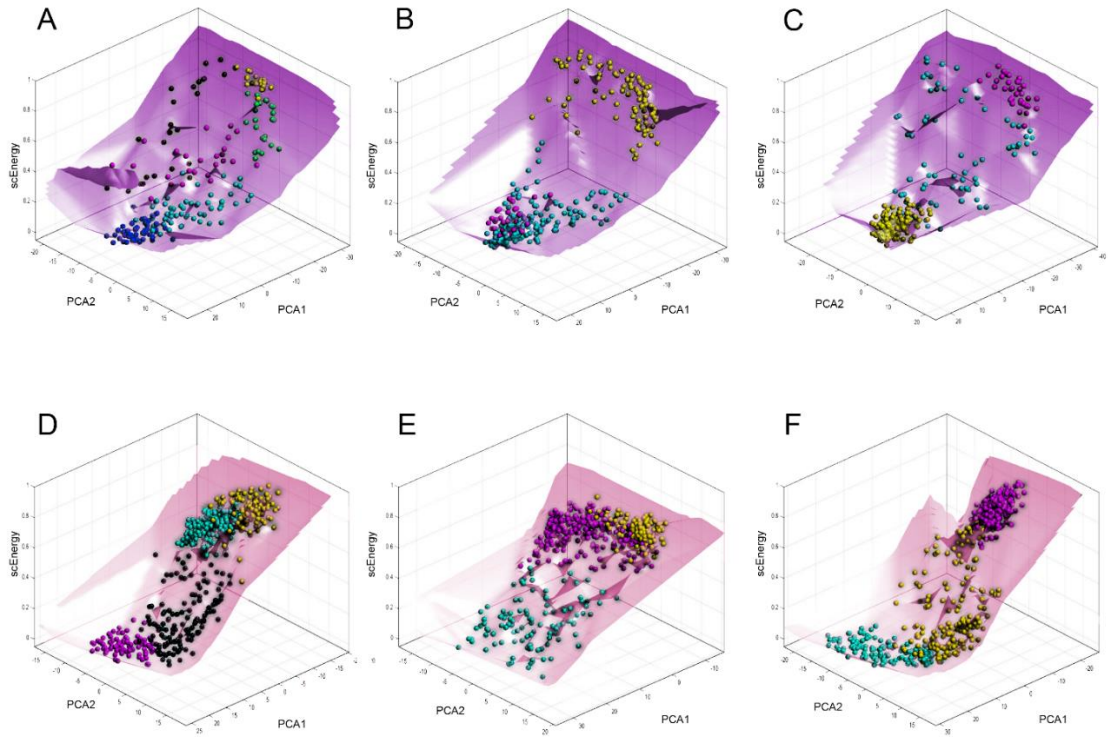

**FIGURE 1. WADDINGTON ENERGY LANDSCAPES OF GBM AND GSC CELLS. A-C)** The scEpath Waddington landscapes of pediatric GBM, and D-F) Waddington landscapes of adult GSC cell fate dynamics. They are color-coded in violet and pink, respectively. Cell states (balls on the landscape) with similar gene expression as determined by an unsupervised clustering framework used by scEpath and the scEnergy calculation (Equation 1) are clustered together within the same color. The horizontal plane axes correspond to the reduced dimensionality space of gene expression determined by principal component analysis (PCA), wherein PCA1 and PCA2 are the first two PCA components. The vertical axis is the scEnergy. The patterns of cell clusters visualized on the Waddington landscape are *attractors*. While distinct attractors may exist on the landscape corresponding to a higher and lower energy state, the global cell fate patterns (i.e., trajectories of cell clusters) on the landscape were treated as a single dynamical structure.

### 3.2. *GAN-generated patterns infer the presence of a strange attractor steering GBM/GSC cell fate dynamics.*

The GAN-inferred attractors all exhibit a flap of the manifold (state-space) being folded at the corner of the attractor, followed by twisting, stretching, and folding to the observed self-similar patterns in both patient groups (Fig 2A-F). The Rössler attractor had a box-count FD of 1.64. The GAN-generated patterns anatomically resemble the Rössler attractor. The Box-count FD of GAN-generated attractors from the cell fate patterns of pediatric GBM was determined to be 1.73 ± 0.04, while that of adult GSC was 1.71 ± 0.02 (Figure 2G). The error bars correspond to the standard deviations as computed by GraphPad Prism. The relatively low error bars demonstrate the precision of the calculated FD. Considering the uncertainties (error bars), the FD of GAN-generated attractors of both patient groups are in good agreement with each other, suggesting the presence of a strange attractor governing the cell fate trajectories of both, pediatric GBM and adult GSC, on the Waddington landscape.

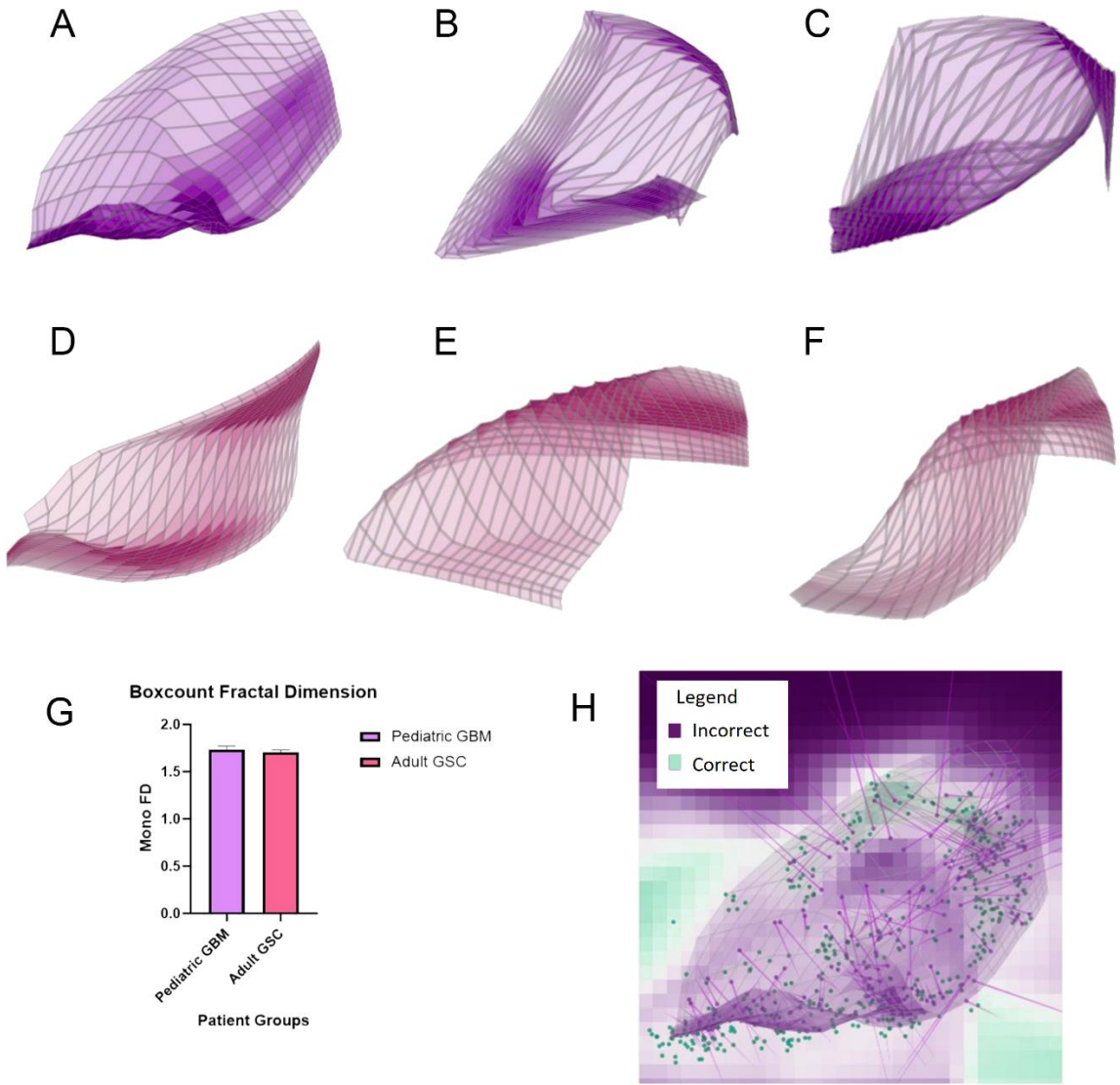

**FIGURE 2. ANATOMY AND FRACTAL DIMENSION OF GAN-GENERATED ATTRACTORS.** A- C) The emergence of attractors in GAN generated patterns are shown for three pediatric GBM samples color-coded in violet. D- F) GAN-generated patterns of three adult GSC samples color coded in pink. The grey lines are the grid lines of a spatial manifold the generator uses to learn and reconstruct an inferred pattern from the input dataset. G) The mono-FD scores from the box-count algorithm on the GAN-generated attractors of both patient groups. Error bars correspond to standard deviations in FD scores. H) A sample image of how the GAN generates patterns from the input dataset. The green dots represent the input data (i.e., the cell state patterns from the scEpath landscapes in Figure 1). Background colors of grid cells represent the discriminator's classification. Turquoise grid cells are likely to be real input data points, while those in purple are likely incorrect (newly generated) (opacity encodes density). When the GAN converges to a newly generated pattern, the maximum amount of grid cells will be purple as shown here. The manifold in purple represents the generator's discovered pattern. The loss metrics, and Kullback-Leibler and Jensen-Shannon divergence metrics were displayed on the GAN-interface to assess the GAN's performance (not shown). The loss metric reported the loss values of the discriminator and generator, while the divergence metrics assessed how similar the input data (real) distributions are from the newly generated patterns (Kahng et al., 2019).

## 4. DISCUSSION

The findings suggest the presence of a Rössler-like strange attractor steering GSC/GBM cell fate dynamics in signaling state-space. The presented study demonstrates the first-time use of Generative Adversarial Networks and Waddington landscape reconstruction as novel complex systems approaches for inferring causal patterns in the complex dynamics driving GBM/GSC evolution. Further, the findings strongly suggest for the first-time a causal link (attractor) between pediatric GBM cells and adult GSCs. The study is a proof-of-concept in the applicability of complex systems physics in advancing healthcare and oncology.

### *4.1. GAN-generated patterns demonstrate the strong plausible existence of a global attractor governing both GBM and GSC cell fate dynamics on gene expression state-space (Waddington landscape).*

GANs are demonstrated as promising Deep learning platforms for attractor reconstruction from GBM patient-derived gene expression datasets. The GAN is not trained to reproduce data but rather generate new data patterns and trick the discriminator it is real (Kahng et al., 2019). As shown in Figure 2A-F, a similar pattern was generated from twelve different GBM and GSC landscapes' cell patterning distribution. As such, the detected causal pattern could plausibly model the dynamics of the complex system (i.e., GBM and GSC cell fate trajectories). Furthermore, these findings suggest the existence of a complex attractor driving GBM/GSC cell fate decisions and disease progression.

### *4.2. Topological signatures of the Rössler attractor were observed in the anatomy of GAN-reconstructed GBM/GSC cell fate patterns.*

Rössler predicted the emergence of a strange attractor, referred to as the Rössler attractor, in biochemical oscillations (Rössler, 1976). The folding stretch as shown in Figure 3 is a characteristic signature observed in the birth of the Rössler attractor. This characteristic flap folding (of phase-space) was detected as a hallmark signature in all twelve GAN-generated patterns. The flap fold was essential within the initial few iterations (epochs) of generating the GAN-patterns. The GAN-reconstructed attractors had a fractal dimension of roughly 1.7 in all n= 12 patient samples (Figure 2G). The emergence of self-similar GAN- reconstructed attractors with equivalent fractal dimensions in all patient samples suggests the presence of a global *strange attractor* driving the complex dynamics of GBM/GSC cells. Further, the computed fractal dimension of all GAN-reconstructed attractors was found to be close in proximity to that of the Rössler attractor.

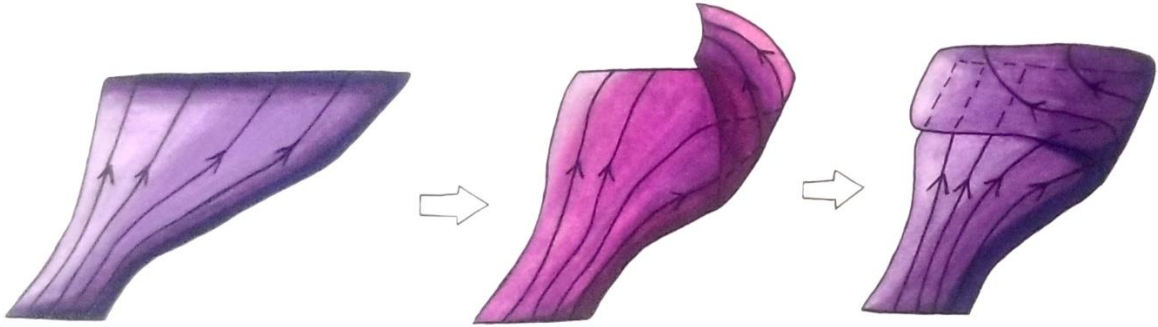

**FIGURE 3. DISSECTION OF THE RÖSSLER ATTRACTOR.** Within the context of a cooking analogy, the birth of a strange attractor resembles the kneading of a puff pastry. Think of phase space as a pastry dough. The dough must be stretched, folded, and repeated for the emergence of flaky, layered pastries. These puff pastries of phase-space steering the trajectories of the complex dynamical system is known as a strange attractor and is characterized by a fractal dimension (Ruelle, 1980). The anatomical cross-section of the state-space flow resembles Smale's horseshoe map (Strogatz, 2015). The presence of a strange attractor is a hallmark of chaotic dynamics within the system (Strogatz, 2015).  [Image was re-designed based on a lecture by Kartofelev, D. (Tallinn University of Technology). Similar reproductions are found in Abraham and Shaw (1983) and Strogatz (2015)].

The emergence of a strange attractor suggests a causal mechanism for the emergence of complex adaptive behaviors such as resilience, phenotypic plasticity/heterogeneity, GBM aggressivity, therapy resistance, and GSC stemness/recurrence (Baish and Jain, 1998; Itik and Banks, 2010; Heltberg et al., 2019). A higher fractal dimension indicates the dynamics are more complex and irregular within the system generating the pattern (Baish and Jain, 1998; Strogatz, 2015; Uthamacumaran, 2021). As such, the calculated fractal dimension of 1.7 can be interpreted as a measure of GBM (cancer) complexity. Furthermore, the findings suggest that the dynamical structure and hence, behaviors of pediatric GBM cells resemble closely that of adult GSCs. A major problem in clinical oncology remains the classification of pediatric GBM into distinct molecular subtypes. Unlike adult GBM, they do not have well defined molecular subgroups as given by the Verhaak classification (Verhaak, 2010; Paugh et al., 2010; Jones et al., 2017). Furthermore, pediatric GBM exhibit a greater epigenetic burden with characteristic histone modifications and DNA methylation profiles, indicating more complex interactions (interdependence) with their tumor microenvironment (Lulla et al., 2016). The findings presented herein suggest that this may be because pediatric GBM cells have closer resemblance to adult cancer stem cells (GSCs) than their mature counterpart, i.e., the differentiated adult GBM cells.

### 4.3. The presence of a strange attractor in GBM/GSC cell fate trajectories provides a causal pattern to screen for robust clinical targets and predict therapeutic responses.

These findings suggest the cell fate dynamics of GBM/GSC cells may be governed by a strange attractor in their signaling state-space. GANs are demonstrated as robust artificial intelligence algorithms for reconstructing complex attractors from signaling state-space and help predict causal patterns in cell fate decision-making. To validate the predictions, systems scientists should perform perturbation analysis on essential drivers of GSC stemness by gene-expression editing techniques (e.g., CRISPR, viral transfection, pharmacological interventions) to observe which perturbation results in the destabilization of the GAN-reconstructed attractor in GBM/GSC scRNA-Seq datasets. The destabilization of the attractor can be assessed by a decrease in its fractal dimension.

*Clinically- relevant propositions:* We can begin by targeting established signaling networks conferring GSC stemness in GBM (e.g., SALL2, OLIG1/2, SOX2, POU3F2, as established by Suvà et al. (2014)), or embryonic morphogens regulating the stem cell niche. The scEpath algorithm used for the Waddington landscape reconstruction provides heat map analyses of gene expression signatures predicted in the cell fate transitions of GBM developmental trajectories. These are additional critical gene targets for such perturbation analysis. The presented blueprint of complex systems approaches can then be used to determine whether the observed GAN-generated attractor reduces (or increases) in fractal dimension. As such, we can identify critical signals required for the GBM/GSC dynamics and predict the evolutionary trajectories of GBM cell fate decisions. Once such genes (transcription factors) are identified, traditional attractor reconstruction methods such as Takens's time-delay coordinate embedding should be assessed on the time-traces of the unperturbed genes' signals to verify if a strange attractor emerges.

*Limitations:* scRNA-Seq datasets of pediatric GSCs were not available. Moreover, we can optimize the GANs pattern analysis/attractor reconstruction from cancer gene expression by training the GAN on data distributions of strange attractors. Further studies should be extended on different stages of GBM disease progression, and other cancer subtypes. Time-series analysis should be paved in prospective studies for verification of the inferred attractors. Despite these limitations, the presented results are a proof-of-concept in the applicability of complex systems theory in predicting the dynamics of complex diseases like cancer. Nonetheless, the presented findings suggest a causal pattern in the form of a strange attractor may be steering GBM cell fate decisions in signaling state-space.

**FUNDING**: N/A

**CONFLICT OF INTEREST**: There are no competing interests.

**AUTHOR CONTRIBUTION:** AU conceived, conducted the analyses, and wrote the manuscript.

**ACKNOWLEDGEMENTS**: Thanks to Dr. Mario D'Amico, Dr. Ingo Salzmann, and Dr. Laszlo Kalman of Concordia University (Dept. of Physics) for mentoring me and help revise/edit the paper. Thanks to Rik Bhattacharja for designing the palette of Figures 1 and 2. Thanks to Dalia Alkayal for designing Figure 3.

**DATA AVAILABILITY AND CODES:** The single-cell datasets (expression matrices) are available in the following repositories. Columns are cell barcodes, row names are genes, in all expression (count) matrices.

*Pediatric GBM:* https://singlecell.broadinstitute.org/single_cell/study/SCP393/single-cell-rna-seq-of-adult-and-pediatric-glioblastoma#study-summary (Neftel et al., 2019)
*Expression Matrix* (Expression Matrix [log2(TPM/10 + 1)] (Smartseq2): IDHwtGBM.processed.SS2.logTPM.txt.gz

*Adult GSC:* https://singlecell.broadinstitute.org/single_cell/study/SCP503 (Richards et al., 2021)
*Count matrix:* Richards_NatureCancer_GSC_scRNAseq_counts.csv.gz

**scEpath algorithm:** https://github.com/sqjin/scEpath (Jin et al., 2018)

**GAN lab:** https://poloclub.github.io/ganlab/ (Kahng et al., 2019)

**FracLac ImageJ Plugin (v2.5)** (Box-Count algorithm): http://rsb.info.nih.gov/ij/plugins/fraclac/FLHelp/Introduction.htm.